\documentclass[rapids]{jfm}
\usepackage{graphicx}
\usepackage{epstopdf,epsfig}

\usepackage{xcolor}
\usepackage{latexsym}
\definecolor{gr}{rgb}{0.5,0.5,0.5}

\shorttitle{On the eddies of the log law}
\shortauthor{M. Heisel, C. M. de Silva, N. Hutchins, I. Marusic and M. Guala}

\title{On the mixing length eddies and logarithmic mean velocity profile in wall turbulence}

\author{Michael Heisel\aff{1,2}
  \corresp{\email{heise070@umn.edu}},
  Charitha M. de Silva\aff{3},
  Nicholas Hutchins\aff{4},
  Ivan Marusic\aff{4},
 \and Michele Guala\aff{1,2}}

\affiliation{\aff{1}St. Anthony Falls Laboratory, University of Minnesota, Minneapolis, MN 55414, USA
\aff{2}Department of Civil, Environmental, and Geo- Engineering, University of Minnesota, Minneapolis, MN 55455, USA
\aff{3}School of Mechanical and Manufacturing Engineering, University of New South Wales, Sydney 2052, Australia
\aff{4}Department of Mechanical Engineering, University of Melbourne, Victoria 3010, Australia}

\begin{document}

\maketitle

\begin{abstract}

Since the introduction of the logarithmic law of the wall more than 80 years ago, the equation for the mean velocity profile in turbulent boundary layers has been widely applied to model near-surface processes and parameterise surface drag. Yet the hypothetical turbulent eddies proposed in the original logarithmic law derivation and mixing length theory of Prandtl have never been conclusively linked to physical features in the flow. Here, we present evidence that suggests these eddies correspond to regions of coherent streamwise momentum known as uniform momentum zones (UMZs). The arrangement of UMZs results in a step-like shape for the instantaneous velocity profile, and the smooth mean profile results from the average UMZ properties, which are shown to scale with the friction velocity and wall-normal distance in the logarithmic region. These findings are confirmed across a wide range of Reynolds number and surface roughness conditions from the laboratory scale to the atmospheric surface layer.

\end{abstract}


\section{Introduction}

Despite the wide-ranging occurrence of turbulent boundary layers in engineering and environmental flows \citep{Schlichting99,Stull88,Brutsaert13} -- and more than one hundred years of research on the subject \citep{Prandtl04} -- experimental and computational constraints remain an obstacle to the advancement of existing theory for high Reynolds number flows. Perhaps the most notable example relates to the logarithmic (log) region of turbulent boundary layers. The mean velocity profile in this region is the subject of ongoing research due to its widespread relevance and impact \citep{Zagarola98,George07,Lvov08,Marusic13}. For instance, the profile can be used to estimate the surface drag on ship hulls and aircrafts \citep{Prandtl55} and is the basis for many empirical relationships in atmospheric and climate applications such as the Monin-Obhukov similarity theory \citep{Monin54}. The mean streamwise velocity $U$ in the log region is described by the so-called log law of the wall, first derived by Ludwig \citet{Prandtl25} and Theodore \citet{Karman30}:

\begin{equation}
U^+ = \frac{1}{\kappa} \mathrm{ln} \left( z^+ \right) + A.
\label{eq1}
\end{equation}

Here, $z$ is the wall-normal position, $\kappa \approx 0.4$ is the von K\'{a}rm\'{a}n constant, and the parameter $A$ depends on the surface roughness. The superscript ``$+$'' indicates normalization in wall units, i.e.  $U^+ = U/u_{\tau}$ and $z^+ = zu_{\tau}/\nu$, where $u_{\tau}$ is the friction velocity corresponding to the average wall shear stress and $\nu$ is the kinematic viscosity. The log region is sufficiently far from both the wall and the outer boundary condition such that the effects of the viscous length scale $\nu/u_{\tau}$ and boundary layer thickness $\delta$ are small, and the primary length scale is the wall-normal distance $z$.  Only recently have laboratory and field facilities reached sufficiently high friction Reynolds number $\Rey_{\tau} = \delta u_{\tau} / \nu$ under well-controlled conditions to rigorously evaluate the universality of equation (\ref{eq1}) \citep{Marusic13}.

Even though the log law of the wall has been supported experimentally, certain underlying assumptions still remain open-ended. The earliest derivations of equation (\ref{eq1}) by \citet{Prandtl25} and \citet{Karman30} rely on an assumed relationship between the shear stress $\tau$ and the mean velocity gradient:

\begin{equation}
\tau = \rho {l_e}^2 \left( \frac{\partial U}{\partial z} \right)^2.
\label{eq2}
\end{equation}

Here, $\rho$ is the fluid density and $l_e$ is the so-called mixing length corresponding to the size of hypothetical ``eddies'' responsible for momentum transfer \citep{Prandtl25}. Taking Prandtl's common form of the mixing length $\ell_e = \kappa z$ for the self-similar log region \citep{Prandtl32}, equation (\ref{eq2}) leads to the mean shear scaling $\partial U / \partial z = u_{\tau} / \kappa z$ whose integral is equation (\ref{eq1}). \citet{Townsend76} described these eddies as turbulent motions ``attached'' to the wall in the sense that their spatial coherence extends to the wall and the average eddy size increases with distance from the wall. Many recent studies have identified wall-attached behavior in the logarithmic region using various analysis methods. Examples include detection of regions with coherent velocity fluctuations \citep{Hwang18} and intense Reynolds stresses \citep{Lozano12,Jimenez18}, simulations filtered to specific length scales \citep{Hwang15}, modal decomposition \citep{Cheng19}, and resolvent analysis \citep{McKeon19}. However, both the mixing length and attached eddies have not yet been unambiguously linked to coherent structures across a range of flow conditions, particularly for rough surfaces and in very-high-Reynolds-number boundary layers.

Studies of coherent structures in high-Reynolds-number flows have revealed the presence of spatial regions with relatively uniform streamwise velocity \citep{Meinhart95}. These structures are known as uniform momentum zones (UMZs) and have been identified both at the laboratory scale \citep{Adrian00,deSilva16,Saxton2017,Laskari18} and in the atmosphere \citep{Morris07,Heisel18}. Individual UMZs are separated by relatively thin regions where a large percentage of the overall shear and vorticity are concentrated \citep{Priyadarshana07,Eisma15,deSilva17}. The approximation of boundary layer turbulence as a series of UMZs is consistent with the mean momentum balance and wall-normal distance scaling of the log region \citep{Klewicki09,Klewicki13}. Due to the arrangement of UMZs and thin vortical regions, the instantaneous velocity profiles resemble a step-like function, and the smooth logarithmic profile is only achieved through long-term averaging \citep{deSilva16}. This framework has been applied to a UMZ--vortical-fissure model for boundary layers \citep{Bautista19}, but the predicted UMZ properties employed by the model have not been confirmed experimentally. Furthermore, previous UMZ studies focused primarily on the outer region in smooth-wall flows and have not provided detailed size statistics and conclusive scaling for the log region.

Accordingly, the present work evaluates the properties of UMZs across a variety of flow conditions, with an emphasis on high-Reynolds-number, zero-pressure-gradient boundary layers. The goal of this work is to reconcile the structural composition of UMZs with mixing length and attached eddies and the derivation of the log law of the wall. The analysis is focused within the log region above the viscous (or roughness) sublayer and below $z \lesssim 0.15 \delta$ \citep{Marusic13}. By using a numerical simulation and an atmospheric flow in addition to laboratory measurements, UMZs are compared across a uniquely large range of both Reynolds number $\Rey_{\tau} \sim O(10^3-10^6)$ and surface roughness $k_s^+ \sim O(0-10^4)$, where $k_s$ is the equivalent sand grain roughness with $k_s^+ \gtrsim 70$ indicating fully rough conditions \citep{Jimenez04}. The orders-of-magnitude differences in $\Rey_{\tau}$ and $k_s^+$ allow for a careful evaluation of the universal scaling behavior of UMZs in both smooth- and rough-wall boundary layers.

\section{Methodology}

\begin{table}
\begin{center}
\def~{\hphantom{0}}
\begin{tabular}{rccccl}
Dataset     					& Label	& Symbol						& $\Rey_{\tau}$	& $k_s^+$	& Source				\\
\hline
direct numerical simulation	& DNS	& \textcolor{gr}{$*$}		& 2\,000        & --			& \citet{Sillero13}	\\
smooth wall    				& sw1	& \textcolor{gr}{$\times$}	& 3\,800        & --			& new              	\\
smooth wall    				& sw2	& \textcolor{gr}{$+$}		& 4\,700        & --			& new           		\\
smooth wall    				& sw3	& \textcolor{gr}{$\bigcirc$}	& 6\,600        & --			& \citet{deSilva14}	\\
smooth wall    				& sw4	& \textcolor{gr}{$\square$}	& 12\,000       & --			& \citet{deSilva14}	\\
smooth wall    				& sw5	& \textcolor{gr}{$\lozenge$}	& 17\,000       & --			& \citet{deSilva14}	\\
mesh roughness      			& m1		& $\bigtriangleup$			& 10\,100       & 430		& new           		\\
mesh roughness      			& m2		& $\bigtriangledown$			& 13\,900       & 620		& new           		\\
sandpaper roughness			& sp1	& $\rhd$						& 12\,000       & 64			& \citet{Squire16}	\\
sandpaper roughness			& sp2	& $\lhd$						& 18\,000       & 104		& \citet{Squire16}	\\
atmospheric surface layer	& ASL	& $\bullet$					& $O(10^6)$     & 30\,000	& \citet{Heisel18}	\\
\end{tabular}
\caption{Experimental datasets used in the comparison of uniform momentum zone (UMZ) properties.}
\label{tbl1}
\end{center}
\end{table}

\subsection{Previous experiments}

Streamwise velocity measurements were collected from seven previously published boundary layer experiments under approximately zero-pressure-gradient conditions. The experiments are summarized in table \ref{tbl1}. The lowest $\Rey_{\tau}$ case is from the direct numerical simulation (DNS) of \citet{Sillero13}. Two-dimensional slices in the streamwise--wall-normal plane were extracted from the DNS results to match the measurement plane of the remaining particle image velocimetry (PIV) experiments. Three of the smooth-wall and the two sandpaper roughness cases are based on large-field-of-view PIV measurements from the High Reynolds Number Boundary Layer Wind Tunnel at the University of Melbourne, which were previously published by \citet{deSilva13,deSilva14} and \citet{Squire16}. The highest $\Rey_{\tau}$ case is from recent super-large-scale PIV measurements in the canonical log region of the atmospheric surface layer (ASL) by \citet{Heisel18}. The ASL measurements represent a practical field setting with very-high Reynolds number in near-neutral thermal stability conditions. Further details on the measurements can be found in the references included in table \ref{tbl1}.

\subsection{New experiments}

To complement the existing databases, new PIV measurements were acquired for two smooth wall and two woven wire mesh roughness cases in the boundary layer wind tunnel at St. Anthony Falls Laboratory. The test section of the closed-loop wind tunnel is 16 m downstream of the contraction and has cross-sectional dimensions of 1.7 $\times$ 1.7 m$^2$ under approximately zero-pressure-gradient conditions. For the rough-wall cases, the test section and fetch were covered with woven wire mesh \citep[see, e.g.,][]{Flack07}. The mesh had 3 mm wire diameter and 25 mm opening size, i.e. distance between wires, resulting in equivalent sand grain roughness $k_s=$ 17 mm. Hotwire anemometer measurements of the full boundary layer profile were used to estimate flow parameters such as $\delta$.

\begin{figure}
\centerline{\includegraphics{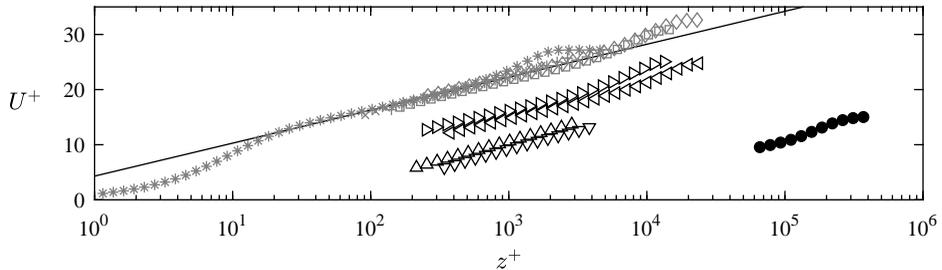}}
\caption{Mean streamwise velocity profiles normalized in wall units. Data symbols correspond to the experiments in table \ref{tbl1} and the line is the log law for smooth wall conditions. Data symbols are shown with logarithmic spacing for clarity.}
\label{fig1}
\end{figure}

The PIV setup in the tunnel test section included a Big Sky 532 nm Nd:YAG double-pulsed laser oriented in the streamwise--wall-normal plane, a TSI Powerview 4 MP camera, and TSI Insight 4G synchronizer and acquisition software. The field of view was limited to the lowest 25\% of the boundary layer in the rough-wall case where $\delta \approx$ 400 mm (50\% in the smooth-wall case where $\delta \approx$ 200 mm) to enhance the spatial resolution in the logarithmic region. In-house cross-correlation code was used to compute the velocity vectors from the images \citep{Nemes15}. The interrogation window size ranged from 25 to 55 wall units depending on the flow conditions. Wall-normal profiles of the mean velocity for the new and previous experiments are shown in figure \ref{fig1}.

\subsection{Detection of UMZs}

For each dataset, UMZs were detected from histograms of the instantaneous $u$ velocity fields \citep{Adrian00}. The histogram method, summarized here, has been proven to successfully identify the organization of the flow into relatively uniform flow regions separated by thin layers of high shear; the variability of $u$ within the detected UMZs is a small fraction of the overall time-averaged variance, and a majority of the instantaneous shear $\partial u / \partial z$ and spanwise vortices are aligned with the identified UMZ interfaces \citep{deSilva16,deSilva17,Heisel18}.

\begin{figure}
\centerline{\includegraphics{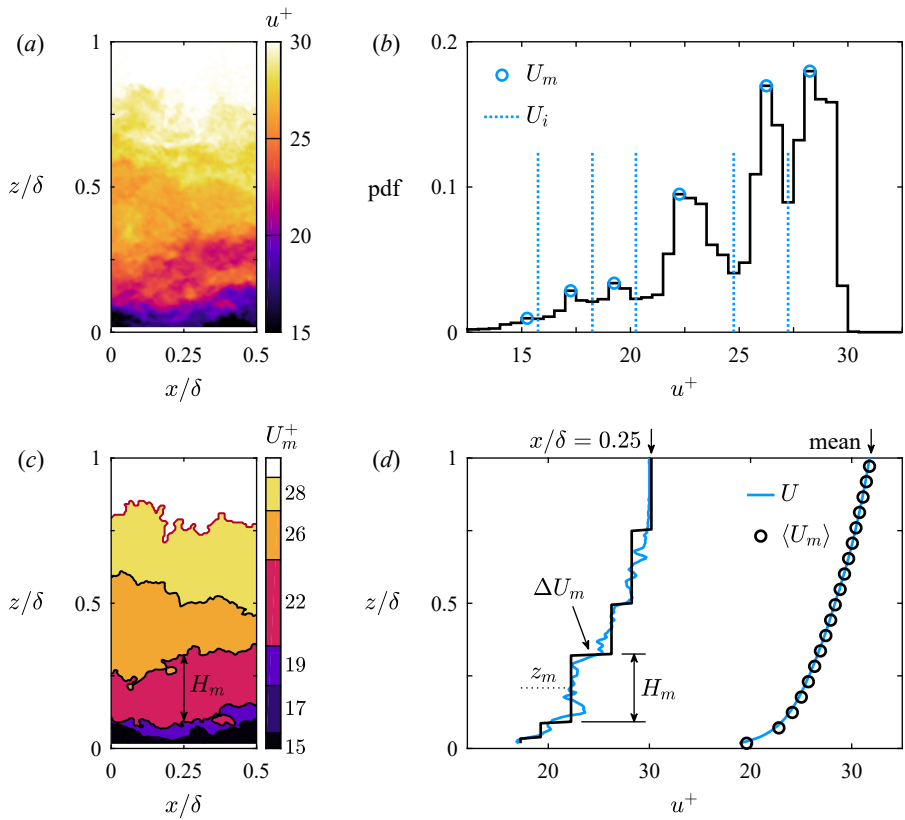}}
\caption{Example detection of UMZs from experiment ``sw5'' in table \ref{tbl1}. (\textit{a}) Streamwise velocity field $u(x,z)$. (\textit{b}) Histogram of the vectors in (\textit{a}) with the detected modes $U_m$ and minima $U_i$. (\textit{c}) Estimated UMZ field including internal UMZ interfaces corresponding to $U_i$ (black lines) and the turbulent/non-turbulent interface (red line). (\textit{d}) Instantaneous and time-averaged profiles of $u$ (blue lines) and $U_m$ (black lines).}
\label{fig2}
\end{figure}

Figure \ref{fig2}(\textit{a,b}) shows an example vector field and histogram, where the free-stream region above the turbulent/non-turbulent interface (TNTI) was excluded from the histogram \citep{deSilva16}. The TNTI was detected using a threshold of the local kinetic energy \citep{Chauhan14}. In the histogram, each mode $U_m$ represents the velocity of a distinct UMZ \citep{Adrian00}. Because the shear interfaces between UMZs are characterized by a large velocity gradient across a short distance, their velocity $U_i$ is represented by a small number of vectors and can be approximated as the minima between modes in the histogram. We hereafter refer to these shear regions as UMZ interfaces, noting that different terminology is used depending on the study \citep[see, e.g.,][]{Priyadarshana07,Eisma15}. The positions associated with the interface velocity were determined using isocontours of the velocity $U_i$, e.g. the black lines separating each UMZ in figure \ref{fig2}(\textit{c}).

The shear interfaces scale in size with the Taylor microscale \citep{Eisma15,deSilva17}. In the logarithmic region of our laboratory-scale datasets, the interfaces cover up to approximately 15\% of the measurement area, and the interior of UMZs covers the remaining 85\% (the UMZ coverage increases weakly with $Re_{\tau}$). Here, we consider the UMZ and its corresponding shear interfaces to be a single unit that defines the representative eddy. The thickness of the UMZ interface and interior are combined in a single parameter $H_m$ which is the wall-normal distance between the center of adjacent interfaces. Combining the thicknesses does not affect the conclusions of the study. The thickness values $H_m$ were compiled for each column in every PIV frame, resulting in at least $10^6$ $H_m$ values for every dataset. To evaluate the thickness as a function of the wall-normal distance, $H_m$ was ensemble averaged in intervals of $z/\delta$, where the UMZ mid-height $z_m$ was used to determine the $z/\delta$ interval for each UMZ.

The method for calculating the velocity difference $\Delta U_m$ across UMZ interfaces requires conditional averaging and hence does not allow for instantaneous estimates \citep{deSilva17,Heisel18}. Wall-normal profiles of the velocity relative to the interface were compiled for all interfaces in every column and frame. Based on the interface wall-normal position, the profiles were sorted using the same $z/\delta$ intervals as for $H_m$. For each interval, the interface profiles were ensemble averaged and $\Delta U_m$ was computed using linear fits to the average profile as detailed in \citet{deSilva17}.

Figure \ref{fig2}(\textit{d}) illustrates the UMZ properties described above. These properties characterize the two-dimensional realization of each three-dimensional UMZ structure as it crosses the measurement plane. Because we cannot assess the spanwise properties of UMZs with the PIV measurements, we focus our analysis on how an ensemble of the two-dimensional realizations relates to the mean velocity profile in the same measurement plane. In this regard, figure \ref{fig2}(\textit{d}) shows close agreement between the wall-normal profile of the mean measured velocity $U$ and the ensemble averaged modal velocity $U_m$.

While figure \ref{fig2} shows a wide velocity field for visualization purposes, in the analysis each field was segmented into sections with streamwise length $\mathcal{L}_x = 0.1 \delta$ prior to the histogram calculation. The lowest wall-normal position reported in later results is $z = 0.05\delta$, and the smallest observed structures are expected to occur at this position. The aspect ratio of the streamwise extent and wall-normal position of wall-attached structures is between 10 and 15 \citep{Baars17} such that the average length of structures at $z = 0.05\delta$ is approximately $0.5\delta$. The length $\mathcal{L}_x = 0.1 \delta$ was selected to be short enough for smaller-than-average structures to manifest distinct peaks in the histograms, while also large enough to yield statistically converged histograms in the dataset with the coarsest resolution.

\begin{figure}
\centerline{\includegraphics{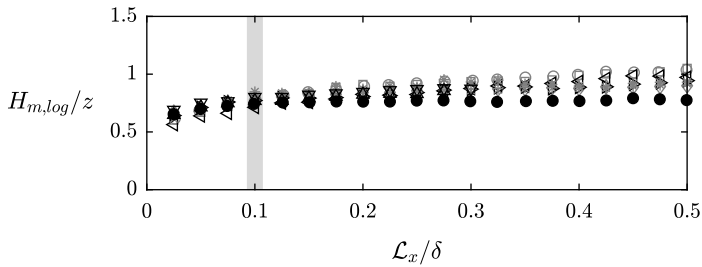}}
\caption{Average UMZ thickness $H_{m,log}$ in the logarithmic region as a function of the detection parameter $\mathcal{L}_x$, where $\mathcal{L}_x=0.1\delta$ is the value used for later results. Data symbols correspond to the experiments in table \ref{tbl1}.}
\label{fig3}
\end{figure}

To assess the sensitivity of the results to the choice of $\mathcal{L}_x$, UMZ thickness statistics were computed for a range of $\mathcal{L}_x$ using a small sample of each dataset. Figure \ref{fig3} plots the average UMZ thickness $H_{m,log}$ within the logarithmic region as a function of $\mathcal{L}_x$. The thickness $H_m$ increases moderately with $\mathcal{L}_x$ due to the exclusion of the smallest structures. The difference between the ASL and lab-scale datasets for larger $\mathcal{L}_x$ may be due to the underestimated size of the largest UMZs exceeding the field of view in the ASL measurements \citep{Heisel18}. Considering the orders-of-magnitude difference in $H_m$, the agreement of results across datasets is minimally affected and the conclusions drawn from the results do not change within the range of $\mathcal{L}_x$ shown.  Figure \ref{fig3} therefore shows $\mathcal{L}_x = 0.1\delta$ to be appropriate for studying $H_m$ in the range of $z/\delta$ presented here.

\section{Results}

\begin{figure}
\centerline{\includegraphics{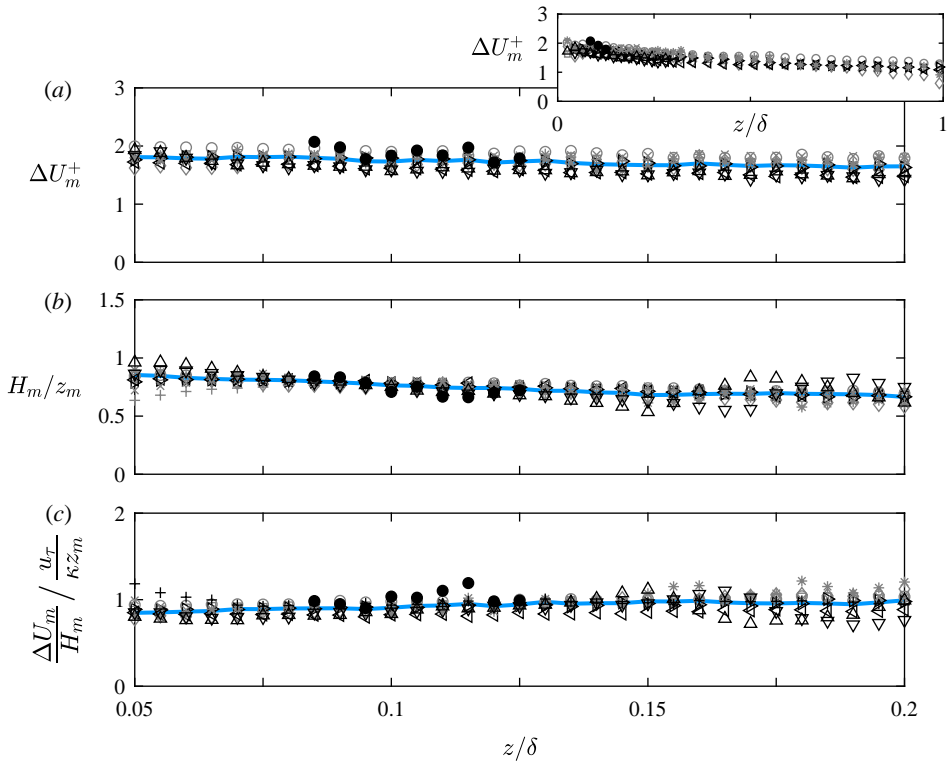}}
\caption{Profiles of average UMZ properties in the logarithmic region. (\textit{a}) Velocity jump $\Delta U_m$ across UMZ interfaces. The inset plot shows $\Delta U_m$ across the entire boundary layer thickness. (\textit{b}) Wall-normal thickness $H_m$ of UMZs. (\textit{c}) Comparison of UMZ properties with the mean shear scaling $u_{\tau} / \kappa z$. Data symbols correspond to the experiments in table \ref{tbl1} and the blue line is the average.}
\label{fig4}
\end{figure}

Figure \ref{fig4} shows wall-normal profiles of the UMZ properties. The approximately continuous ensemble-averaged UMZ profiles result from variability in the UMZ size, velocity, and position throughout the averaging period. The stochastic behaviour of the interface position was studied in \citet{deSilva17}, and variability in the UMZ size is addressed later in this section. As seen in figure \ref{fig4}(\textit{a}) and the inset plot, the UMZ characteristic velocity $\Delta U_m$ scales unambiguously with the friction velocity $u_{\tau}$ throughout the entire boundary layer. The result agrees with the log law formulation and highlights $u_{\tau}$ as the relevant turbulent velocity scale across the entire boundary layer \citep{Smits11}. The decreasing trend with increasing $z$ can be explained by viscous diffusion of the gradients with increasing distance from the wall \citep{Tsinober01}. The observed moderate decrease from $\Delta U_m \approx 1.7 u_{\tau}$ in the log region to $\Delta U_m \approx u_{\tau}$ near the edge of the boundary layer compliments previous laboratory-scale results \citep{deSilva17}, and the present work extends the $u_{\tau}$ scaling to a wider range of $\Rey_{\tau}$ and surface roughness.

The UMZ thickness $H_m$ in figure \ref{fig4}(\textit{b}) shows the wall-normal distance to be the appropriate length scale for UMZs in the log region. To emphasize the success of the normalization, note that the dimensional range represented in figure \ref{fig4}(\textit{b}) is $H_m=$ 0.0065 to 6.3 m. Based on the observed functional dependence $H_m \sim z_m$ in this region, the UMZs exhibit the wall-attached behavior predicted by the mixing length. Any differences between the smooth- and rough-wall profiles is within the uncertainty of the results.

Figure \ref{fig4}(\textit{a,b}) highlights the similarity of UMZs relative to the log law scaling parameters $u_{\tau}$ and $z$ across a three order-of-magnitude range in Reynolds number. In figure \ref{fig4}(\textit{c}), the mean shear $\partial U / \partial z$ is estimated from the average UMZ properties $\Delta U_m / H_m$ and is compared with the shear scaling argument $u_\tau / \kappa z$. The values near unity demonstrate the fundamental interdependence of the UMZ properties and the shear scaling parameters in the log region, where the distribution of UMZs relates directly to the mean velocity gradient as $\Delta U_m / H_m \approx u_{\tau} / \kappa z$. Consequently, figure \ref{fig4}(\textit{c}) suggests the UMZs to be a good structural model for recovering mean velocity behavior in the log region for high Reynolds number flows. In other words, UMZs correspond to the turbulent motions associated with the mean shear, therefore providing a link between the attached eddies in a statistical sense and physical features in the turbulent flow.

Taking the UMZ and its associated shear interface to be the representative attached eddy, the UMZ properties can be compared with the previously discussed mixing length predictions. The predictions consider the normalized eddy velocity $u_e/u_{\tau}$ and size $\ell_e/z$ to be constant in the log region. However, our results in figure \ref{fig4}(\textit{a,b}) suggest $u_e/u_{\tau}$ and $\ell_e/z$ may both be weakly dependent on the wall-normal position. Additionally, the average values within the log region $\Delta U_m \approx 1.7 u_{\tau}$ and $H_m \approx 0.75 z$ are approximately twice the mixing length theory values $u_e = u_{\tau}$ and $\ell_e = \kappa z$ \citep{Prandtl32}, and are closer to more recent predictions $u_e = 1.62 u_{\tau}$ and $\ell_e = 0.62 z$ \citep{Bautista19}. Our results for $\Delta U_m$ and $H_m$, confirmed unambiguously across a range of $\Rey_{\tau}$ and $k_s$, provide new values for the velocity and length of the eddies leading to the mean shear. As a note of caution, the definition of UMZs does not explicitly include wall-normal advection, and $H_m$ is not necessarily equivalent to the length scale for wall-normal momentum transport.

\begin{figure}
\centerline{\includegraphics{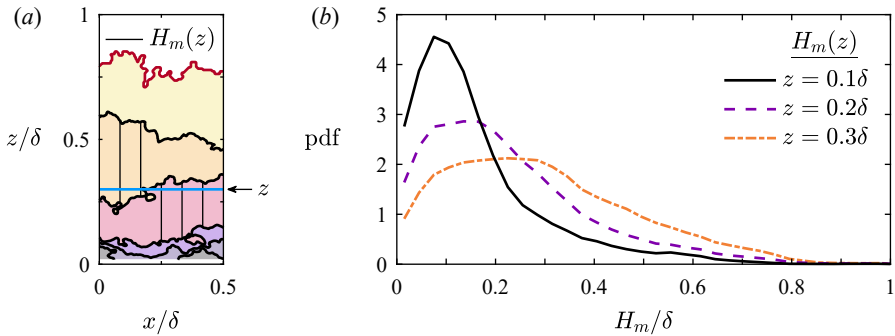}}
\caption{Example compilation of UMZ thickness statistics for the probability analysis. (\textit{a}) The statistics at a given position $z$ (blue line) include every thickness $H_m$ where the UMZ intersects with $z$. (\textit{b}) Probability densities of $H_m(z)$ at three wall-normal positions for the DNS dataset.}
\label{fig5}
\end{figure}

Thus far, the observed ``attached'' behavior has been based on the functional dependence $H_m \sim z_m$ of the \textit{average} UMZ thickness. The influence of $z$ and other scaling parameters such as $\delta$ on individual UMZs can be investigated through probability distributions of the UMZ size. The previous shear scaling comparison in figure \ref{fig4}(\textit{c}) required the use of a single representative height for each UMZ, i.e. the midheight $H_m(z_m)$. In a probability analysis, however, the influence of large UMZs extending to the near-wall region would be diminished by confining the UMZ to a single height $z_m$ farther from the wall. To avoid this bias, we calculate the height-dependent probability statistics using a new selection criterion on the original $H_m$ values. At a given wall-normal position $z$, the thickness statistic $H_m(z)$ includes every UMZ which reaches the position $z$ as depicted in figure \ref{fig5}(\textit{a}). The statistics are repeated for all $z$. Examples of $H_m(z)$ probability density functions (pdfs) at three $z$ positions for the DNS dataset are shown in figure \ref{fig5}(\textit{b}).

\begin{figure}
\centerline{\includegraphics{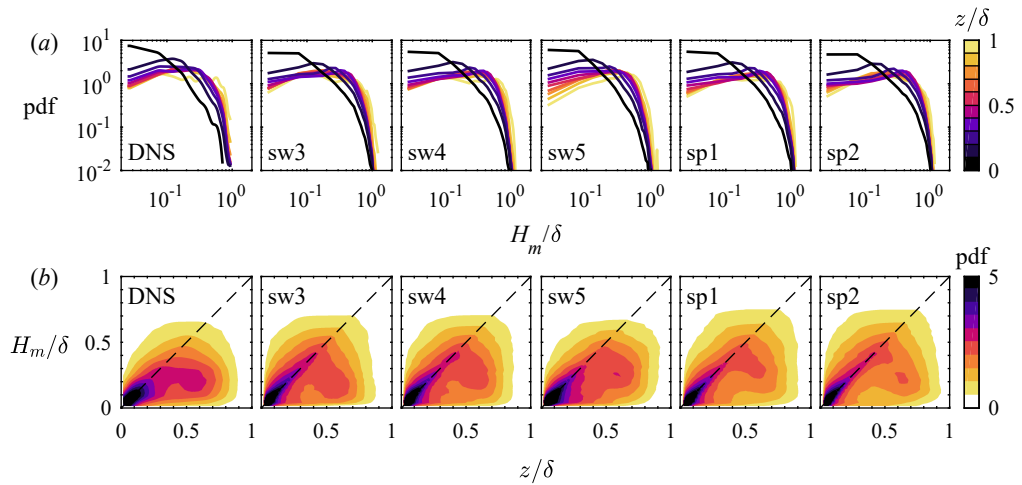}}
\caption{Probability density functions (pdfs) of UMZ thickness $H_m(z)$ for every dataset where the field of view extended to $z=\delta$. (\textit{a}) Separate pdfs for different wall-normal positions $z$ indicated by the line color. (\textit{b}) Joint pdfs of $H_m$ and $z$, where the dashed line represents $H_m(z) = z$. Columns correspond to the indicated experiments.}
\label{fig6}
\end{figure}

Figure \ref{fig6} shows the resulting probability distributions of $H_m(z)$ for the datasets where the field of view included the full boundary layer thickness. The plots in figure \ref{fig6}(\textit{a}) suggest that the distribution tail is limited by $H_m \lesssim \delta$ regardless of position $z$, which is expected. However, for small $z / \delta$ in the log region, the largest structures limited by the outer condition $\delta$ are rarely occurring such that the influence of $\delta$ on the mean behavior is small. As the wall-normal distance increases, $\delta$ limits the size of an increasing proportion of the identified structures. The behavior of the probability tail explains the decreasing trend in figure \ref{fig4}(\textit{b}), where $H_m / z$ decreases slowly with $z$ as $\delta$ becomes increasingly relevant. The joint probabilities of $H_m$ and $z$ in figure \ref{fig6}(\textit{b}) more clearly shows the departure from wall-attached behavior in the log region to outer ($\delta$) scaling in the wake region. The small differences between cases are likely due to differences in the experimental measurement resolution and variability in the detection process. The consistent trend across datasets is that the most probable UMZ thickness follows $H_m(z) = z$ up to $z / \delta \approx 0.5$, above which the probability distribution transitions to being independent of wall-normal distance as $\delta$ becomes the primary length scale.

\section{Concluding remarks}

In this work we built on previous studies that approximate high-Reynolds-number turbulent boundary layers as a series of uniform flow regions and associated shear interfaces. This intermittent flow organization has also been suggested for homogeneous turbulence \citep[e.g., see][]{Ishihara13,Elsinga17}. In this approximation, the low-amplitude turbulence within the UMZs is neglected. Here we also assume the velocity difference $\Delta U_m$ occurs instantaneously as shown in figure \ref{fig2}(\textit{d}), as opposed to occurring across the Taylor microscale-thickness of the shear interface. This assumption does not affect the shear results in figure \ref{fig4}(\textit{c}). The purpose of the UMZ approximation as seen in figure \ref{fig2}(\textit{c}) is to demonstrate the importance of these prominent layered structures to the mean velocity statistics. While each instantaneous flow field has a discrete number of UMZs, variability in the UMZ position and velocity over space and time leads to the seemingly continuous average properties in figure \ref{fig4}.

The experimental results -- specifically the collapse of the profiles in figure \ref{fig4} -- show the scaling behavior of the UMZs to be universal for zero-pressure-gradient boundary layers regardless of Reynolds number and surface roughness, including atmospheric flows in fully rough conditions. The independence of Reynolds number in the results supports the assumption of complete similarity in the derivation of equation (\ref{eq1}) and is consistent with Townsend's outer layer similarity hypothesis (\citeyear{Townsend76}). Within the log region, UMZ properties are governed by the theoretical scaling parameters $u_{\tau}$ and $z$. The findings provide experimental evidence that the hypothetical eddies assumed in early log law derivations have a well-defined physical representation in the layered structure of wall turbulence. The UMZs and the eddies of Prandtl’s mixing length model both result in the same mean shear scaling and $\kappa$ value leading to the log law of the wall. The difference between the UMZ size $H_m \approx 0.75 z$ and the common mixing length definition $\ell_e = \kappa z$ is not surprising considering the von K\'{a}rm\'{a}n constant $\kappa$ originates from mean velocity measurements. The results provide a reminder that the coefficient $\kappa$ is specifically associated with the ratio $u_{\tau}/z$ and it therefore does not quantify the precise size of energy-containing turbulent motions as is assumed by the definition $\ell_e = \kappa z$.

In addition to the theoretical implications, the results demonstrate the feasibility of reduced-order modeling of high Reynolds boundary layer flows, whereby streamwise velocity characteristics can be reproduced by modeling only the coherent uniform momentum regions and the vorticity in their interfaces to create the step-like instantaneous velocity profiles. This framework is adopted by the UMZ--vortical-fissure model \citep{Bautista19} and is closely related to the velocity fields resulting from the attached eddy model \citep{Marusic19,deSilva16}. The present work confirms the scaling relationships assumed in these models, and provides benchmark statistics and probability distributions which show similarity in the zonal organization across a wide range of flow and surface conditions. Moreover, the extension of the UMZ structural similarity to the ASL shows the modeling approach to be viable for atmospheric flows under near-neutral conditions, thus demonstrating the opportunity for simplified near-surface flow models in applications ranging from micrometeorology to wind energy.

% Acknowledgments
The authors acknowledge funding support from the Institute on the Environment (IonE) and the Australian Research Council. M.G. is supported by a National Science Foundation CAREER grant (NSF-CBET-1351303). M.H.'s visit to the University of Melbourne was supported by the University of Minnesota Graduate School. The authors are grateful to J.A. Sillero for allowing public access to the DNS data.

\bibliographystyle{jfm}
\bibliography{JFM-19-RP-1587_references}

\end{document}